\documentclass[conference]{IEEEtran}
\usepackage[utf8x]{inputenc}

\ifCLASSINFOpdf
  \usepackage[pdftex]{graphicx}
  
\else
  
\fi

\usepackage{amsfonts}
\usepackage{amsmath}
\usepackage{amssymb}
\usepackage{amssymb}
\usepackage{amsthm,epsfig,float}

\author{
    \IEEEauthorblockN{Prakash Chaki, Gouri Nawathe, Aaqib Patel, S.N. Merchant, U.B. Desai}
    \IEEEauthorblockA{Department of Electrical Engineering\\Indian Institute of Technology, Bombay\\Mumbai, India 400 076
    \\\{prakashc,gouri,aaqib,merchant,ubdesai\}@ee.iitb.ac.in}

}

\title{Symbiotic Cognitive Relaying with mobile Secondary nodes in Cognitive Radio Networks}

\begin{document}

\maketitle

\begin{abstract}
In a Symbiotic Cognitive Relaying (SCR) scenario, Secondary User (SU) nodes can act as multihop relays to assist communication between Primary User (PU) nodes in the
case of a weak direct link. In return, the SU nodes are incentivised with the right to carry out SU-SU communication 
using licensed PU band for a fixed amount of time, referred to as the `Time Incentive'. Existing work on SCR is constrained to a fixed ad-hoc SU network. In this paper, we 
introduce mobility in SCR by considering mobile SU nodes while keeping the PU nodes fixed. 
This paper uses a specific mobility pattern and routing strategy for the SU nodes to propose theoretical bounds on the throughput and delay for PU-PU 
transmission. We derive analytically the least throughput and maximum delay possible in our model.
\end{abstract}


\IEEEpeerreviewmaketitle

\section{Introduction}
Communication today has been posed with the problem of overcrowding in unlicensed bands, a direct consequence of the static spectrum allocation policy. Cognitive Radios offer an exciting 
solution to this problem by intelligently exploiting the under-utilized spectrum in licensed bands without causing any hindrance to
the licensed user’s communication. Co-operative communication is an interesting paradigm in which Cognitive Radios have the potential of creating tremendous 
impact. In its simplest form, co-operative communication consists of an intermediate node which relays data between 
the transmitter and receiver, located far away from each other, in case of a weak direct communication link between them. This results in an improvement in the link quality and in turn, the 
system capacity. Communication using multiple hop relays not only helps in overcoming path loss and achieve power gain, but 
also adds multipath diversity, critical to wireless communication over fading channels[8]. T. Nadkar $\textit{et al.}$
have shown in [1] that in an SCR scenario, SU nodes can act as multihop
relays to maximize throughput between the PU transmitter and the PU receiver, which have a very weak direct communication link between them. The time which the PU saves through the 
SCR approach is rewarded to the SU nodes to carry out SU-SU communication using
licensed PU band. Similar to this model, termed as \textit{`Cognitive Relaying with time incentive'}[1], the authors in [1] also propose another 
scheme with frequency usage incentives. In this model, while relaying the PU's data, SU nodes only try to achieve the throughput that the primary would have 
achieved on its own on the direct weak link, but SUs require a lesser number of channels to achieve the same. The remaining frequency channels are awarded to the SUs for their own 
communication. Authors in [6] have discussed an interesting \textit{cooperative cognitive radio network}[6] 
where the primary transmitter asks for relay-based cooperation from some other PU or SU node for its transmission to the primary 
receiver. This drastically improves the throughput of primary communication and the transmission is completed at a higher data rate.
In return, the SU nodes may be allowed to use the licensed primary band for the saved time duration by paying some revenue. 
Authors in [6] formulate a Stackelberg game between the PU(as leader) and SU nodes(as follower) where the PU nodes try to maximize its throughput and revenue using the SU nodes 
while the SU nodes tries to use the channel at the least cost. Simeone \emph{et al.} in [7] also takes the \textit{property-rights model} 
of spectrum-usage and discuss a model where the PU node chooses a subset of SU nodes and allows them to use primary band for some 
units of time asking for co-operation by relaying of primary's data in return. In the awarded time, the SU nodes compete with each 
other for their own communication by power control in a distributed fashion[7]. Thus, all these literatures motivates us to have a detailed look into the 
\textit{Property-rights model} in contrast to the much studied \textit{Commons model} and see if the primary can leverage its 
spectrum to the secondary nodes asking for a better QoS for itself. Mobility in Cognitive Radio networks also hasn't been a much studied area. Mobility brings in notable impact on various 
parameters of a CR network including but not limited to spectrum opportunity, handoff, PU protection range, spatio-temporal diversity, sensing scheduling[9] etc. In this work, by introducing
 mobility to the SU nodes following a specific mobility pattern, while keeping the PU nodes stationary, we analyze the throughput in each of the relaying phases and also find a bound on the delay 
for SCR model in Cognitive Radio network.\\
The rest of this paper is organized as follows. In Sec. II, we discuss the related works on theoretical bounds of throughput in mobile networks. Sec. III talks about our system model. In Sec. IV, 
we show the detailed throughput analysis and we conclude the paper with results and discussions in Sec. V.  

\section{Related Work}

Gupta and Kumar[2] have shown in their pioneering work on capacity of ad-hoc wireless networks that the per node average 
achievable throughput for a ad-hoc network with randomly distributed static nodes is $\Theta(\frac{W}{\sqrt{nlogn}})$ in a non-interfering 
scenario where W is the available bandwidth to the network. Even by optimally choosing the positions of the nodes and
 their transmission ranges, the best average per-node throughput that can be achieved is $\Theta(\frac{W}{\sqrt{n}})$. The results in [1] also 
assumes an optimal scheduler which knows the location and traffic patterns of all the nodes. If they are not known or 
the nodes starts moving, then the capacity will be smaller than this.  In other words, the maximum bit-meters* per second 
that can be supported by an ad-hoc wireless network on a disk of unit area is $\Theta(\frac{W}{\sqrt{n}})$. Thus, the authors suggests 
considering networks with smaller no. of nodes or scenario where each node communicates with its nearest neighbour 
since increase in node density n reduces the capacity. Tse and Grossglauser[3] shows that by introducing mobility and 
using a 2-hop relay algorithm in between each S-D pair, the per node throughput can be increased from $\Theta(\frac{1}{\sqrt{nlogn}})$ or 
$\Theta(\frac{1}{\sqrt{n}})$ to $\Theta(n)$. This surprising increase in throughput owes its reason to the mobility of nodes, scheduling policy and 
the buffering and relaying of data at the intermediate nodes.  The main result in [3] is that even though n increases, 
the long-term throughput per source(S)-destination(D) pair remains almost unaffected. But the authors in [3] don’t 
consider the delay caused in such a model. The results in [3] are mostly useful for systems which can afford to incur
 a large delay to increase the throughput. The key problem in a fixed ad-hoc network is that direct communication from
 S to D becomes extremely difficult when the no. of nodes increases[3]. So, intermediate nodes are used as relays.
 Usually, a multi-hop route consists of nodes of the order $\sqrt{n}$. So, when n is very high, the nodes are always busy with 
relaying. Thus, effective throughput of the network comes down. The solution to reduce number of hops is sought by introducing 
mobility. The nodes transmit only when they come close to the destination. But this can’t be a realistic solution. Authors
 in [3] suggests a routing algorithm by which a source node distributes its packets to its nearest neighbors. These mobile 
relay nodes then pass on the packet to the final destination when any of them go close to the destination. As n increases, 
the probability of at least one of the mobile relays going close to the destination becomes very high. Thus, a source node 
uses the multiple user diversity to transfer its packet to the destination node. Bansal and Liu shows in [4] that by 
following a specific routing algorithm, the per node throughput can be \emph{c}.$\frac{W.min(m,n)}{nlog^{3}n}$ keeping the delay less than $\frac{2d}{v}$.

\section{System Model}

In our network model, we consider $n$ static Primary User (PU) and \emph{m} Secondary User (SU) nodes. The PU nodes are static and infrastructure-based.
The SU nodes, on the other hand, are mobile which move around following a specific mobility model that we discuss next. We assume that every node knows each other's position 
at every instant of time by using some mechanism like GPS or centralized location database. The link between the PU $T_X$ and PU $R_X$ 
being very weak as considered in [1], the PU nodes seek help from the mobile SU nodes to relay their packets at a higher data rate. The SU nodes in turn 
get \emph{time-incentive}[1] to carry out their own communication on the licensed PU band without any need of spectrum sensing and thus they save their own energy as well. 
The wireless channel that we consider in our setup is in the same spirit as of most of the other works[2]-[5], i.e, we consider large scale path loss and ignore the small scale 
multipath fading. So, channel gain is defined as 
\begin{gather*}
 g(h)=h^{-\gamma}
\end{gather*}where \emph{h} is the distance between the source(S)-destination(D) pair and $\gamma$ is the path-loss factor. We call the transmission from a source 
node \emph{i} to a destination node \emph{j} at time \emph{t} is successful if  
\begin{gather*}
\frac{P_i(t)g_{ij}(t)}{N_0 + \sum \limits_{k\neq i}P_i(t)g_{ij}(t)} > \zeta
\end{gather*}
where, $P_{i}(t)$ is the transmission power of node \emph{i} at time \emph{t} and $\zeta$ is the threshold Signal to Noise Ratio required for a successful transmission. 

\subsection{Mobility pattern of SU nodes}

We define the mobility pattern of the SU nodes as follows:\\
The initial positions of the SU nodes are randomly chosen from a uniform distribution. Then, these SU nodes start moving choosing velocity and direction in the 
following manner:\\
{\bf Velocity:} All the SU nodes moves with a same fixed velocity $v$ throughout the time.\\
{\bf Direction:} The initial direction of motion for all SU nodes are \emph{i.i.d} and generated at random from (0,2$\pi$]. Each SU node moves in this 
direction for a pre-defined time interval \emph{t}. After moving in a particular direction for time \emph{t}, the nodes changes its direction to a new direction
 which lies in an angular range $\pm \frac{\alpha}{2}$ from its previous direction. This way of modelling the direction of mobile nodes has been done keeping in view the very 
realistic situations where a moving node (for eg, a wireless device handheld by a person or fixed to a vehicle) usually moves in a specific direction and doesn't 
change abruptly to any new random direction which is poles apart from its previous direction.

\begin{figure}[ht] 
\centering 
\label{con} 
\includegraphics[scale=0.32]{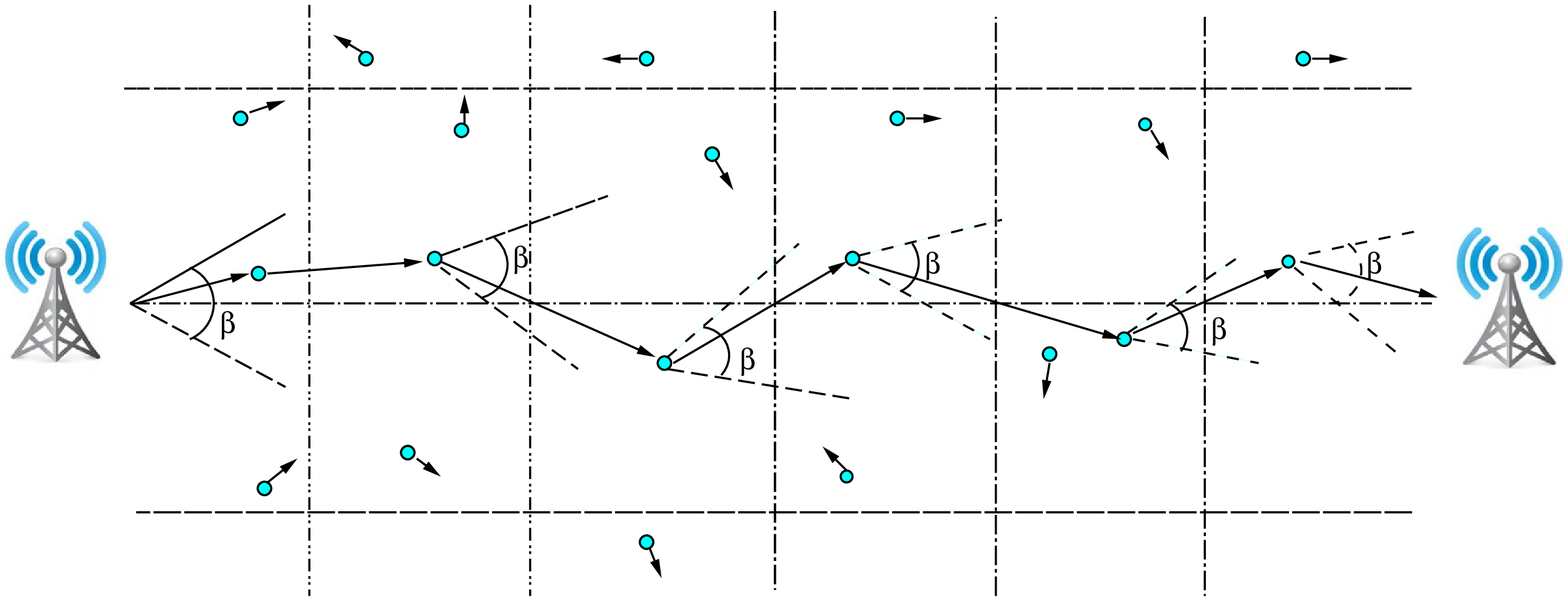}\\
\caption{Network model showing PU $T_X$-$R_X$ pair and mobile SU nodes.} 
\vspace{-2.25mm}
\end{figure}

\subsection{Transmission and relaying model:} The PU $T_X$ firstly looks at the straight line joining itself with the PU $R_X$, and scans in the angular range $\pm \frac{\beta}{2}$ for a nearby SU node (see fig.1). 
We also define two boundary lines parallel to and at a distance $d$ from the straight line joining the PU $T_X$ and PU $R_X$. Relaying of packets is done in the following 
3 phase scheme as it has been done in [3],[4]: PU $T_X$-SU relaying, SU-SU relaying and SU-PU $R_X$ relaying. But we propose an entirely new model of routing the packets to increase
throughput.

\emph{Case(a): PU $T_X$-SU relaying}\\
In every time slot, PU $T_X$ selects one SU node for relaying its packet based on the following parameters in decreasing order of priorities:
\begin{enumerate}
 \item \textit{Direction of motion}: We define two parallel lines at a distance \emph{d} from the straight line joining the PU $T_X$ and PU $R_X$ as shown in Fig.1. The PU $T_X$ 
selects a SU node for relaying which lies in between these two 
straight lines and moves towards the line joining the PU $T_X$ and PU $R_X$ and also towards the PU $R_X$. The objective is to have a minimum path length from the PU $T_X$ 
to the PU $R_X$, so the SU nodes moving closely around the line joining the PU $T_X$ and PU $R_X$ are preferred.

\item \textit{Closeness to the PU $T_X$}: A SU node lying close to the PU $T_X$ is preferred over a far-away node as nearest neighbour communication requires less transmission power and thus
reduces interference and increases throughput as shown in [2].

\end{enumerate}

\emph{Case(b): SU-SU relaying}\\ In this phase of relaying also, the parameters considered to select a node is same as above. A SU node looks in the angular range $\pm \frac{\beta}{2}$ 
from the straight line joining itself to the PU $R_X$ in search of a close-by SU node to relay the PU's packet.\\ \\
\emph{Case(c): SU-PU $R_X$ relaying}\\ In this last phase of the multihop communication, when the SU node finds that it has come close to the PU $R_X$, it passes on the packet to the PU $R_X$.

\section{Throughput Analysis}
\subsection{Static PU to Mobile SU relaying phase}
The static PU transmitter node looks for a useful mobile node (useful node is one which satisfies all the conditions mentioned above) in the angular range $\pm \frac{\beta}{2}$ from the 
straight line joining itself to the PU receiver and transfers its packets. We now show a bound on the closeness factor between the PU and SU so that the communication is always successful. 
Theorem 3 of [4] specifies the least condition under which the communication between static PU node and mobile SU node is always a success. We use this result to find a circular 
region around the PU transmitter so that the PU-SU communication is always successful inside this region. We also include the proof of Theorem 3 of [4] here for easy reference.\\

\emph{\bf Theorem:} The packet transmission from a static PU to a mobile SU node is always successful if the SU node enters a circle of radius $\frac{k}{\sqrt{2m}}$ which has the PU node 
at its centre.

\emph{Proof:} From Theorem 3 of [4], assume a square region of side $l=\frac{k}{\sqrt{m}}$ with the static PU transmitter lying at the centre(0,0) of the square. The maximum distance at which a 
mobile SU can lie from the PU transmitter inside this square is $d=\frac{l}{\sqrt{2}}$. Assuming that the PU transmits at unit power and the the path-loss factor $\alpha>$2, received power at the 
SU node from PU is $\frac{1}{d^\alpha}$, which is equal to $\frac{\sqrt{(2m)^\alpha}}{k^\alpha}$. Interference caused at our SU node from the other PU transmitter nodes lying in the successive 
tiers of the cellular region is given as
\begin{gather*}
\leq \sum_{j=1}^{\infty} \frac{8j}{(j-\frac{1}{2})^\alpha m^{-\frac{\alpha}{2}}} \quad \quad \quad \quad \quad \quad\quad\quad\quad\quad\quad\quad[4]\\
=k_1m^{\frac{\alpha}{2}}\quad \quad \quad \quad \quad \quad\quad\quad\quad\quad\quad \quad \quad \quad\quad \quad
\end{gather*}
\vspace{4mm}
where $k_1$ is some constant. Considering $N_0$ as the noise power spectral density, SINR$\geq\frac{\frac{\sqrt{(2m)^\alpha}}{k^\alpha}}{N_0+k_1m^{\frac{\alpha}{2}}}$

So, $k$ can be now chosen to make the SINR greater than the threshold value for successful communication.\\
We now rotate this square at various angles keeping its centre fixed to (0,0). The locus of the diagonal points thus gives us a circle of radius $d$ which also satisfies the above 
theorem by symmetry.\quad\quad\quad\quad\quad\quad\quad\quad\quad\quad\quad\quad\quad\quad\quad\quad\quad\quad\quad\quad$\blacksquare$\\
We now intend to find a bound on how long the static PU transmitter has to wait to find an \textit{`useful'} SU node to relay its data in the lines of Lemma 4 in [4].

\textit{\bf Claim:} The time for which the static PU transmitter has to wait till a mobile SU enters its region of successful communication is bounded as follows:\\
\begin{gather*}
\mathbb{P}\Big(T>\sqrt{\frac{8}{\Phi}f\log m}.\frac{1}{v\sqrt{m}}\Big)\leq m^{-f}
\end{gather*}
$Proof:$ Consider fig. 2, the PU transmitter $P$ is located at the centre of the circle of radius $\frac{k}{\sqrt{2m}}$and a mobile SU node $S$ is lying outside the circle moving towards
$P$ such that $|SP|=vT$. Going 
by our mobility pattern defined earlier, $S$ is free to choose any direction only in the angular range $\pm\frac{\alpha}{2}$ at the beginning of a time-frame. So, the probability 
that $S$ chooses a direction so as to enter the circle is dependent on the direction of its velocity in the previous time-slot. Let us assume that $S$ changes its direction every $T$
time frames. 

\begin{figure}[ht] 
\centering 
\label{con} 
\includegraphics[scale=0.5]{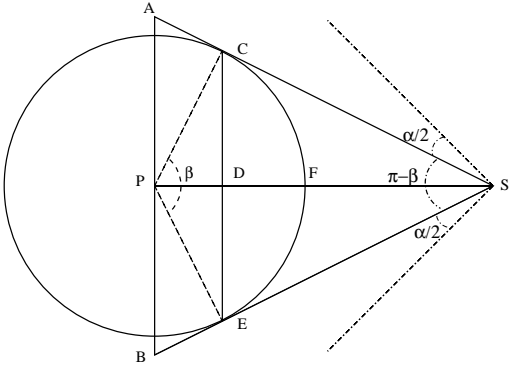}\\
\caption{Static PU to Mobile SU relaying phase.} 
\vspace{0.05in}
\end{figure}

Assuming $t=nT$, current direction of motion
\begin{gather*}
U(t)=U(t-1)+\delta\frac{\alpha}{2}
\end{gather*}
where $\delta$ is a continuous uniform random variable such that $\delta\in[-1,1]$ which is in accordance with our proposed mobility model. Let $U(t=0)=\chi$, then for any $t$,
\begin{align*}
U(t)&=U(t-1)+\delta_1\frac{\alpha}{2}\\
&=U(t-2)+\delta_1\frac{\alpha}{2}+\delta_2\frac{\alpha}{2},\quad\delta_1,\delta_2\in[-1,1]
\end{align*}
Solving iteratively for $U(t)$ we get,
\begin{gather*}
U(t)=\chi+\sum_{i=1}^{n}\delta_i\frac{\alpha}{2} ; \quad\quad\delta_i\in[-1,1]
\end{gather*}

By Central Limit Theorem, for large $n$, $\frac{\sum_{i=1}^{n}\delta_i}{\sqrt{m}}=Z$ where $Z\sim\mathcal{N}(0,\sigma^2)$, ${\sigma^2}=\frac{n\alpha^2}{12}$. So, $U(t)$ is a sum of two independent random variables and its probability density function is thus obtained by 
calculating the convolution of the probability density functions of $\chi$ and $Z$. Hence, $f_U(u)=\displaystyle\int_{-2\pi+u}^{u}\frac{1}{2\pi}(\frac{1}{\sqrt{2\pi\sigma^2}}e^\frac{-z^2}{2\sigma^2})dz=\frac{1}{2\pi}[Q(u-2\pi)-Q(u)]$ .
The communication between node $P$ and node $S$ is successful only if $S$ enters sector ${PCE}$. $S$ will enter sector ${PCE}$ for sure if it chooses a direction within the angular range of $\angle{CSE}$.
Hence, probability of successful communication between nodes $P$ and $S$ is given by $\displaystyle\int_{\frac{-(\pi-\beta)}{2}}^{\frac{(\pi-\beta)}{2}}\frac{1}{2\pi}[Q(u-2\pi)-Q(u)]du$. Using the standard bounds of the $Q$ function (refer to Appendix), we get the probability of successful communication as
\begin{align*}
Pr\geq\ &\displaystyle\int_{\frac{-(\pi-\beta)}{2}}^{\frac{(\pi-\beta)}{2}}\frac{1}{2\pi}\frac{(u-2\pi)e^\frac{-(u-2\pi)^2}{2\sigma^2}}{\sqrt{2\pi}[1+(u-2\pi)^2]}du\\
&-\displaystyle\int_{\frac{-(\pi-\beta)}{2}}^{\frac{(\pi+\beta)}{2}}{\frac{1}{\sqrt{2\pi}}}\frac{e^{\frac{-u^2}{2}}}{u}du
\end{align*}
Since the second term on the right hand side of the inequality is the integral of an odd function, its value will be $0$. The first term on the right hand side is a decreasing 
function, as shown in Fig. 3. The area under consideration is lower bounded by $\square{ABCD}$ which means that $Pr\geq{{e^{\frac{1}{2}}}a\sqrt{2\pi}}\frac{e^{-(\frac{1+(a+2\pi)^2}
{2})}}{1+(a+2\pi)^2}=\Phi(say)$. Here, we have assumed $a=\frac{\pi-\beta}{2}$ for ease of representation.
Now, we consider concentric circles  centred at S. Each of the annular rings have width of $\frac{1}{\sqrt{m}}$. So, there are $\sqrt{m}$ number of rings possible.
Probability of node $M$ lying in the $j^{th}$ annular ring is given by\\$\frac{\mbox{Area of $j^{th}$ ring}}{\mbox{Total area of the disc}}= 
\frac{\frac{\pi}{m}(j^2-(j-1)^2)}{\pi}=\frac{1}{m}(j^2-(j-1)^2)=\frac{2j-1}{m}$\\
Define $i.i.d.$ Bernoulli random variables $N_{ij}$ as follows:\\
$N_{ij}$ is 1 if $i^{th}$ mobile node is in $j^{th}$ annular ring and moving towards the sector $SCHE$, 0 otherwise.
So, $N_{ij}$ indicates if a particular mobile node is \textit{useful} or not in terms of relaying.\\
Hence, $E[N_{ij}]>\Phi\frac{1}{m}(j^2-(j-1)^2)) =\Phi\frac{2j-1}{m}$.
\begin{figure}[H] 
\centering 
\label{con} 
\includegraphics[scale=0.35]{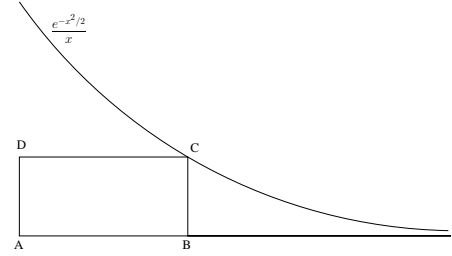}
\caption{Lower bound on area of $\frac{e^{{-x^2}/2}}{x}$}
\end{figure}
Now, counting the number of \textit{useful} nodes within a radius $r$ is
\begin{gather*}
N(r)=\sum_{j=1}^{r}\sum_{i=1}^{m}N_{ij}\\
\therefore E[N(r)]=E[\sum_{j=1}^{r}\sum_{i=1}^{m}N_{ij}]=\sum_{j=1}^{r}\sum_{i=1}^{m}E[N_{ij}]\\>\sum_{j=1}^{r}\sum_{i=1}^{m}\Phi\frac{2j-1}{m}\\
>\Phi\frac{1}{m}r^2m=r^2\Phi
\end{gather*}
By Markov's Inequality, $\mathbb{P}(N(r)\geq1)\leq E[N(r)]$\\
or, $\mathbb{P}(N(r)\geq1)\leq\frac{r^2}{2}\Phi$\\
We obtain a stricter bound using Mutiplicative form of Chernoff's bound as follows :\\
$\mathbb{P}\Big(N(r)\leq(1-\delta)E[N(r)]\Big)<e^{\frac{-\delta^2}{2}E[N(r)]}$ \quad for $0\leq\delta\leq1$\\
Putting $E[N(r)]=\Phi r^2$, $\mathbb{P}\Big(N(r)\leq(1-\delta)\Phi r^2]\Big)<e^{\frac{-\Phi r^2\delta^2}{2}}$\\
Assuming, $\delta=\frac{1}{2}$, $\mathbb{P}\Big(N(r)\leq\frac{\Phi r^2}{2}\Big)<e^{\frac{-\Phi r^2}{8}}$\\
Putting $r^2=\frac{8}{\Phi}f\log m, \mathbb{P}\Big(N(r)<4f\log m\Big)<e^{-f\log m}=m^{-f}.$ Now, as $m\rightarrow\infty, \mathbb{P}\Big(N(r)<4f\log m\Big)\rightarrow0.$\\
Thus, time taken by a mobile node to come close to the PU node is at most $\sqrt{\frac{8}{\Phi}f\log m}.\frac{1}{v\sqrt{m}}\quad\quad\quad\quad\quad\quad\quad\quad\quad\quad\blacksquare$ Now 
if we assume $m^{-f}=0.0001$ i.e.$f=\frac{4\log 10}{\log m}$, we can say that the delay will be greater than $\sqrt{\frac{32\log 10}{\Phi v \sqrt{m}}}$ almost certainly.\\
\emph{\bf Throughput Calculation:}\\
We calculate throughput as:
\vspace{-2mm}
\begin{align*}
\mbox{Throughput}=\ &(\mathbb{E}[\mbox{Data successfully transmitted/interaction])}\\
\ &\times\mbox{(No. of interactions)}
\end{align*}
Expected data successfully transmitted per interaction is at least $\frac{\lambda r_0}{\|v\|}(1-\frac{1}{8\log m})$ by Lemma 6 of [4] where $\lambda$ is the available bandwidth. The minimum 
number of interactions can be calculated as $\frac{Minimum\hspace{1mm}Distance}{Average\hspace{1mm}distance\hspace{1mm}covered\hspace{1mm}between\hspace{1mm}two\hspace{1mm}interactions}$.
\\Minimum number of interactions = $\frac{2R}{{\|v\|}\sqrt{\frac{32log 10}{\Phi{\|v\|} \sqrt{m}}}}$ . Thus,
\begin{align*}
Throughput_{s-m}&\geq\frac{\lambda}{\|v\|}R_0\big(1-\frac{1}{8\log m}\big)\frac{2R}{{\|v\|}\sqrt{\frac{32log 10}{\Phi{\|v\|} \sqrt{m}}}}\\
&=\lambda R_0R \sqrt{\frac{\Phi\sqrt{m}}{8\|v\|^3\log 10}}\big(1-\frac{1}{8\log m}\big)
\end{align*}


\subsection{Mobile SU to Mobile SU relaying phase}
A mobile SU node keeps moving by changing its direction only within a range  $\pm \frac{\alpha}{2}$ from its direction in the previous time slot. It looks for a \textit{useful} node who can assist it in relaying PU's packets to the PU destination. Now, we proved in the static to mobile relaying phase that if a mobile node enters the circle of radius $\frac{k}{\sqrt{2m}}$ with the static PU node at its centre, then such a communication is always successful. Mobile to mobile relaying is possible only when two mobile nodes comes in each other's transmission range as shown below.

In mobile to mobile relaying phase, the transmitting mobile node can move a distance of $x=vt$ in time $t$. So, assuming that the transmitting node can move in any possible direction, the 
circle gets shifted as the centre is translated by a length $x$. So, the intersection of all such circles is now the effective region for successful communication. Radius of this smaller 
circle is essentially $\frac{k}{\sqrt{2m}}-x$. Let us call this radius $R_1$.
\begin{figure}[ht] 
\centering 
\label{con} 
\includegraphics[scale=0.5]{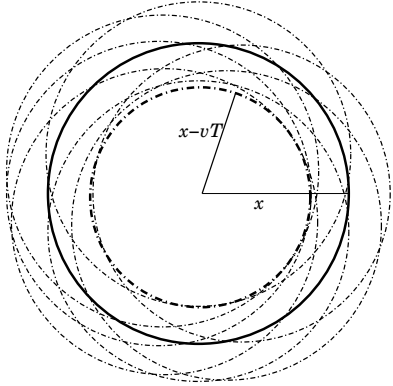}\\
\caption{Successful communication region shrinks to a smaller circle for mobile nodes.} 
\vspace{0.05in}
\end{figure}
Using the result of Lemma 5 of [4], it can be further shown that if two mobile nodes are at a distance less than $\zeta^{-\frac{1}{\alpha}}\frac{1}{\sqrt{8\pi m\log m}}$, then the probability 
of successful communication is greater than or equal to $\frac{1}{8\log m}$ where $\zeta$ is the threshold SINR for successful transmission, $\alpha$ is the path-loss factor and $m$ is the number 
of mobile nodes.\\
\emph{\bf Throughput Calculation:}
\begin{align*}
Throughput_{m-m}&\geq\frac{\lambda}{2\|v\|}R_1\big(1-\frac{1}{8\log m}\big)\frac{2R}{{\|v\|}\sqrt{\frac{32log 10}{\Phi{\|v\|} \sqrt{m}}}}\\
&=\lambda R_0 R \sqrt{\frac{\Phi\sqrt{m}}{32\|v\|^3\log 10}}\big(1-\frac{1}{8\log m}\big)
\end{align*}
\subsection{Mobile SU to static PU relaying phase}
When a mobile SU node finally arrives sufficiently close to the static PU receiver, it finally hands over the primary packet to the PU receiver in this last phase of the relaying process.\\
\emph{\bf Throughput Calculation:}\\
By symmetry, throughput in mobile to static relaying phase is same as obtained in the static to mobile relaying phase. Therefore, 
\\$Throughput_{m-s} \geq 
\lambda R_0 R \sqrt{\frac{\Phi\sqrt{m}}{8\|v\|^3\log 10}}\big(1-\frac{1}{8\log m}\big)$\\
So, net effective throughput for the entire relaying path from static PU transmitter to static PU receiver is $min\big(Throughput_{s-m}, Throughput_{m-m},Throughput_{m-s}\big).$\\
$R_0$ is different for static and mobile phases relaying phases. So, the above expression amounts to minimum of $(R_1,2R_0)$. Since $R_1 = R_0-x$, it is obvious that $R_1<2R_0$. Thus, 
our conclusion is that throughput is greater than $\lambda R_0 R \sqrt{\frac{\Phi\sqrt{m}}{32\|v\|^3\log 10}}\big(1-\frac{1}{8\log m}\big)$. Now, the net delay incurred in transferring 
a single packet from PU transmitter to the PU receiver is obtained as a 
sum of the waiting times for each relay node to find another relay node. Note that we assume that the time required to 
transfer a packet from one node to another is sufficiently small compared to the waiting time for relaying, so it 
doesn't make any difference in delay. We showed earlier that the time that a node has to wait until another node 
satisfying all the relaying criteria comes in its vicinity is $\frac{2R}{{\|v\|}\sqrt{\frac{32log 10}{\Phi{\|v\|} \sqrt{m}}}}$ with probability 0.9999. If a packet is delivered 
in $i$ interactions, the total delay incurred by the packet is $i\sqrt{\frac{8}{k}f\log m}\frac{1}{v\sqrt{m}}$. Since we have assumed that all the mobile nodes travel with the same 
speed $\|v\|$, it is very unlikely that one node would receive the same packet more than once in the course of the packet's journey. Thus, maximum number of interactions is $m+1$.\\
Thus, total delay $\leq\sqrt{\frac{32\log 10}{\Phi\|v\|\sqrt{m}}}(m+1)$.


\section{Conclusion}
Existing literatures in Cognitive Radio doesn't talk much about mobile nodes. In addition to that, most of the work focus on the \textit{Commons Model} of spectrum access and 
\textit{Property Rights Model} is much less explored. We bring these two relatively unexplored areas together in this paper and show analytically the lower and upper theoretical 
bounds on throughput and delay respectively specific to our defined mobility pattern and relaying strategy. One of the basic assumption used in computing the bounds on throughput 
and delay is that the SU nodes moves with a non-zero velocity. 
Since throughput and delay have been calculated in terms of \textit{number of interactions} which is a fixed constant in case of a static network, the obtained results are valid 
only for $v>0$. The throughput value will satisfy the obtained bound with an associated probability of 0.9999 which is fairly high. Taking a closer look at the throughput will 
reveal that it decreases as the number of mobile nodes increases, since an increase in the number of secondary nodes leads to an 
increase in the interference as well, thus enhancing the chances of a packet being dropped. An increase in $\beta$ leads to a decrease in the throughput because more nodes 
would then be eligible to receive packets, which would directly translate in greater interference. Delay on the other hand, decreases drastically with a small increase in the 
number of nodes because waiting time for finding a suitable relay node reduces with an increase in the number of nodes. Similarly, a small 
increase in the speed of mobile nodes leads to a drastic decrease in the total delay, which again can be established intuitively owing to the increase in number of interactions. 
Although Tse and Grossglauser's algorithm[3] guarantees a high throughput, it is more suited to delay-tolerant systems, which is unlike most of the real-time applications.
What is interesting to note is how we can leverage our relay-selection algorithm to increase throughput. A higher throughput can be achieved by 
restricting $\beta$ to small values. A suitable value of $\beta$ could be selected to exploit the throughput-delay tradeoff.
\section{Appendix}
\emph{\bf Lemma:} For static to mobile relaying phase, the expected time for interaction is $\frac{4R_0}{\pi v}$ where $R_0$ is the radius of successful communication region of 
a static node.\\ 
$Proof: $ Consider Fig.3 where $M_0$ is a static PU node
 lying at the centre of the circle of radius $r_0$. $M_i$ be a mobile node which moves at a velocity making an angle $\phi$ to the straight line joining $M_0$ to the PU receiver. Max. distance
that $M_1$ can travel staying inside the circle around $M_0$ is $2R_0$ and their relative velocity is $v\cos\phi$. So, fraction of time for which $M_1$ is in the successful 
communication region around $M_0$ is given as $T=\frac{2R_0}{\|v\cos\phi\|}$. Let, $\psi$ be the angle subtended by the endpoint $D$ of the chord $AD$ at the centre. Now, if $d_i$ is the direction
 of the node $M_1$ and $p_i$ is the angle subtended by the endpoint $D$ of the chord $AD$ at the centre, then 
\begin{align*}
E[T]&=\displaystyle\int\limits_\psi\displaystyle\int\limits_\phi E[T_i|d_i=\phi,p_i=\psi]\mathbb{P}(d_i=\phi,p_i=\psi)\\
&=\displaystyle\int\limits_{-\pi}^\pi\displaystyle\int\limits_\phi^{\frac{\pi}{2}+\phi}\frac{2R_0\cos(\psi-\phi)}{v}\frac{\mathrm{d}\psi}{\frac{\pi}{2}}\frac{\mathrm{d}\phi}{2\pi}=\frac{4R_0}
{\pi v}\quad\quad\quad\quad\quad\blacksquare
\end{align*}
\emph{\bf Lemma:} For mobile to mobile relaying phase, the expected time for interaction is $\frac{16R_0}{v\pi^2}$ where $R_0$ is the radius of successful communication region of a mobile node.
$Proof: $Consider Fig.3, $M_0$ is now moving along the line AB towards B. So, relative distance between $M_0$ and $M_1$ is $2R_0\cos(\psi-\phi)$ and relative 
velocity of $M_1$ w.r.t. $M_0$ is $(v-v\cos\phi)$. Thus, $T=\frac{2R_0\cos(\psi-\phi)}{v-v\cos\phi}\\ E[T]=\displaystyle\int\limits_\psi\displaystyle\int\limits_\phi E[T_i|d_i=\phi,p_i=\psi]\mathbb{P}
(d_i=\phi,p_i=\psi)$\\$=\displaystyle\int\limits_{-\pi}^\pi\displaystyle\int\limits_\phi^{\frac{\pi}{2}+\phi}\frac{2R_0\cos(\psi-\phi)}{v(1-\cos\phi)}\frac{\mathrm{d}\psi}{\frac{\pi}{2}}\frac{\mathrm{d}\phi}{2\pi}
=\frac{2R_0}{v\pi^2}\displaystyle\int\limits_{-\pi}^\pi\frac{1}{(1-\cos\phi)}\mathrm{d}\phi\\=\frac{R_0}{v\pi^2}\displaystyle\int\limits_\pi^\pi cosec^2\frac{\phi}{2}=\frac{16R_0}{v\pi^2}\quad\quad\quad\quad\quad\quad\quad\quad\quad\quad\quad\quad\quad\blacksquare$
\begin{figure}[ht] 
\centering 
\label{con} 
\includegraphics[scale=0.5]{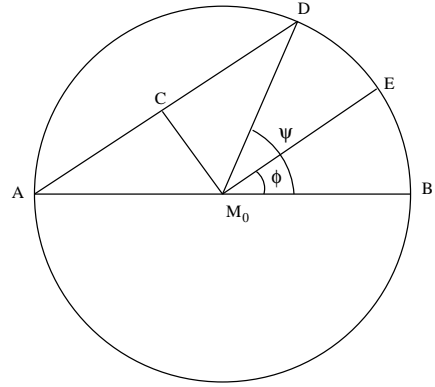}\\
\caption{Calculation of expected time of interaction.} 
\vspace{0.05in}
\end{figure}
\emph{\bf Lemma:} For mobile to static relaying phase, the expected time for interaction is same as that of static to mobile phase.
$Proof: $By symmetry, the expected time for interaction is same as that of static to mobile phase.\quad\quad\quad\quad\quad\quad\quad\quad\quad$\blacksquare$
\emph{\bf Bound on Q(x):}\\$\frac{x}{1+x^2}.\frac{1}{\sqrt{2\pi}}e^\frac{-x^2}{2}<Q(x)<\frac{1}{x}.\frac{1}{\sqrt{2\pi}}e^{-\frac{x^2}{2}},\quad\quad x>0$\quad\quad[10].
\section{Acknowledgement}
This work has been supported by the Ministry of Communication and Information Technology, Government of India, New Delhi.

\end{document}